# Dogfooding: use IBM Cloud services to monitor IBM Cloud infrastructure

William Pourmajidi
Department of Computer Science,
Ryerson University
Toronto, Canada
william.pourmajidi@ryerson.ca

Andriy Miranskyy
Department of Computer Science,
Ryerson University
Toronto, Canada
avm@ryerson.ca

John Steinbacher
IBM Canada Lab
Toronto, Canada
jstein@ca.ibm.com

Tony Erwin
IBM Watson and Cloud Platform
Austin, USA
aerwin@us.ibm.com

David Godwin
IBM Canada Lab
Toronto, Canada
godwin@ca.ibm.com

## ABSTRACT

The stability and performance of Cloud platforms are essential as they directly impact customers' satisfaction. Cloud service providers use Cloud monitoring tools to ensure that rendered services match the quality of service requirements indicated in established contracts such as service-level agreements.

Given the enormous number of resources that need to be monitored, highly scalable and capable monitoring tools are designed and implemented by Cloud service providers such as Amazon, Google, IBM, and Microsoft. Cloud monitoring tools monitor millions of virtual and physical resources and continuously generate logs for each one of them. Considering that logs magnify any technical issue, they can be used for disaster detection, prevention, and recovery. However, logs are useless if they are not assessed and analyzed promptly. Thus, we argue that the scale of Cloud-generated logs makes it impossible for DevOps teams to analyze them effectively. This implies that one needs to automate the process of monitoring and analysis (e.g., using machine learning and artificial intelligence). If the automation will witness an anomaly in the logs — it will alert DevOps staff.

The automatic anomaly detectors require a reliable and scalable platform for gathering, filtering, and transforming the logs, executing the detector models, and sending out the alerts to the DevOps staff. In this work, we report on implementing a prototype of such a platform based on the 7-layered architecture pattern, which leverages micro-service principles to distribute tasks among highly scalable, resources-efficient modules. The modules interact with each other via an instance of the Publish-Subscribe architectural pattern. The platform is deployed on the IBM Cloud service infrastructure and is used to detect anomalies in logs emitted by the IBM Cloud services, hence the dogfooding. In particular, we leverage IBM Cloud Functions to deploy the computing modules, IBM Event Streams to establish communication among the modules, and IBM Cloud Object Storage and IBM Cloudant for persistent storage.

The prototype efficiency is promising: it takes the platform 17 seconds or less from the point of receiving a new log record to emitting an alert to the IBM Cloud DevOps team.

## 1 INTRODUCTION

The use of Cloud computing has become increasingly popular, and both small, and enterprise businesses continuously search for ways to replace their traditional IT practices with services provided by Cloud providers. Ease of use, high level of accessibility, and lower total cost are among promising features that make Cloud offerings both technically and financially attractive compared to traditional, in-house, IT practices. As predicted at the IDC FutureScape in 2018, enterprises will spend more than $530 Billion on Cloud services and Cloud infrastructure by the year 2021 [17].

While Cloud Service Consumers (CSCs) continue to remove barriers to adopt Cloud services, Cloud Service Providers (CSPs) continually invest money and time on enhancing their current offerings and keep on adding new services to their service catalogue. With the advancements in virtualization and Cloud infrastructure management techniques, many CSPs have started offering services that are above traditional computing and storage resources.

In other words, services such as Infrastructure-as-a-Service (IaaS) and Platform-as-a-Service (PaaS) have become ordinary Cloud service models, and both CSPs and CSCs have increased their expectations from Cloud computing to offer services that are often packaged under an umbrella known as Everything-as-a-Service (XaaS) [36]. Services such as Disaster Recovery as a Service (DRaaS), Communications as a Service (CaaS), and Network as a Service (NaaS) are examples of services that are above traditional Cloud computing offerings and can be considered services of Cloud 2.0 [48].

In recent years, the speed of expansion for advanced service delivery has accelerated and has resulted in a new wave of application programming interface (API)-based Cloud offerings that do not see business processes as discrete vertical offerings that are operating in dedicated silos, but rather as a collection of horizontal services that can be accessed across organizational boundaries [31].

The ever-growing demand and dependability of CSCs over the services provided by the CSPs have increased the importance of reliable service delivery for CSPs. For the CSPs, reliability and stability of offered services play a vital role as many services, and their expected quality of service (QoS) are communicated within the service level agreements (SLAs). Any deviation from the promised QoS will negatively impact the reputation of the CSPs, may lead to loss of revenue, and in extreme cases, may become a legal liability for the CSPs [59].

In response to this demand, CSPs have designed and adopted complex Cloud monitoring practices to create a holistic monitoring framework that monitors all elements of Cloud service delivery, from the temperature of a CPU running a hypervisor all the way to the throughput of the network for a disaster recovery data center located on a different continent.

While CSPs mastered the process of generation of the monitoring data, the activities related to gathering, storing, filtering, analyzing, and making prompt decisions based on the monitoring data remain challenging.

In the following subsections, we provide further details about the importance of Cloud monitoring and characteristics of Cloud-generated logs.

## 1.1 Importance of Cloud monitoring

CSPs should constantly monitor the health of their platform to ensure that CSCs are satisfied with the rendered services (meeting quality dimensions: availability, reliability, performance, etc.). Thus, Cloud monitoring is an essential component of a Cloud platform, and it plays a significant role for the CSPs. Similar to Cloud providers, the CSCs who deploy their services on the Cloud require detailed monitoring data and need to keep track of the state of their platform in real-time.

Therefore, the monitoring platform employed by the CSPs should allow them to share a portion of monitoring details with respective CSCs so that each CSC can access a subset of logs and monitoring details that are associated with the components they are using.

The large number of organizations and end-users who are migrating their traditional platforms to Cloud-based platforms, as well as new organizations that begin their business by using XaaS offerings, translate to millions of virtualized elements, each of which requires monitoring and management. In [53], Birje and Bulla acknowledge the rise of the need for implementation better Cloud monitoring platforms and emphasize that Cloud monitoring is necessary for the smooth operation of critical components such as accounting and billing, SLA management, service and resource provisioning, capacity planning, security, and privacy assurance, as well as fault management. Furthermore, the authors indicate that Cloud monitoring tools often consist of five key functions: data collection, data filtering, data aggregation, data analysis, and decision making. In this paper, we propose a solution that can assist CSPs in all of these five key functions.

In [28], Aceto et al. indicate that the number of Cloud-based services has increased rapidly, and as a result, the complexity of infrastructure behind these services has tremendously increased. Therefore, to properly operate and manage a complex infrastructure system, an efficient and effective monitoring system is needed.

## 1.2 Characteristics of Cloud monitoring and Cloud-generated logs

Monitoring solutions are responsible for monitoring various resources in a deployed platform and generate useful insights based on the collected metrics and their respective values. While the majority of monitoring systems are capable of generating graphical reports and sending alerts, the fundamental components of any monitoring system are the components involved in log collection and storage.

Here, we are referring to raw data (generated by each Cloud hardware and software component) and stored for troubleshooting activities. In case of any technical issue, it does not matter which monitoring solution or approach has been used to collect the logs; the actual logs play the most significant role. In fact, Cloud logs are among the most essential pieces of analytical data in Cloud-based services [46]. To effectively manage and offer services, CSPs need to collect and analyze logs to maintain the operation of their platforms at an optimal level.

Moreover, logs are evidential documents [27, 55] and contain all the details and QoS metrics related to the operation of software, network components, servers, and Cloud platforms. As a critical element in computer forensic investigations, logs are presentable in the court of law [58] if they satisfy the legal requirement of admissibility. Logs need to be kept safe and accessible, and a tamper-proof storage system should be used to ensure the authenticity, veracity, and trustworthiness. We are working on such a system [55, 57] to incorporate it into the holistic log processing pipeline.

The rest of this paper is structured as follows. Section 2 describes challenges related to storage and analysis of Cloud-generated logs. Section 3 summarizes the main objective we pursue in this work. Section 4 provides details as to what approaches and components are used to construct the proposed solution. Section 5 reports and discusses the results of implementing the proposed solution using IBM Cloud. Section 6 concludes the paper.

## 2 CHALLENGES

We have described characteristics of the Cloud monitoring and logs in Section 1.2. The log themselves exhibit characteristics of Big Data [44], such as, velocity, volume, value, variety, and veracity [40, 45, 49, 51, 56]. Let us look at some manifestations of these characteristics below.

## 2.1 Challenges of Cloud monitoring

CSPs design and build data centers that host all facilities and equipment that are used to build Cloud infrastructure. It is not unusual for a modern data center to have more than 100,000 components [39] and many network devices such as routers, switches, and firewalls. Naturally, each of these components continuously generates several logs records every second, and as a result, the volume of generated data can be enormous.

To put the size in perspective, let us assume that each of the 100,000 components in a data center generates only one log record per second[1]. It is estimated that 100,000 log records per second will require 2.35 TB of storage space per day even on this 'slow day'[2] [34]. Clearly, the most senior level DevOps engineers can not effectively monitor and analyze 2.35 TB of logs per day.

In addition to the logs generated by the Cloud infrastructure that is mainly to be used by the DevOps staff of the Cloud provider,

---
[1] Obviously, this assumption neglects the fact that each component generates tens of logs per second (hardware metrics, operating system metrics, and application-specific metrics), but, for the sake of argument, let us assume that it is a 'slow day'.
[2] Moreover, the space requirement quickly compounds, as logs may have to be retained for prolonged periods of time. For example, one-year retention would require 858 TB of storage space.

CSPs often need to provide a monitoring platform to the CSCs so they can use it to monitor their Cloud-based deployments. In the case of CSCs with a large number of users, like Netflix, the metrics that need to be monitored produce more than 10 billion records a day [30].

While the high volume (size) of Cloud-generated logs impose a set of challenges related to capacity, storage, and retention of data, the high velocity of Cloud-generated logs increases the pressure on log collection and storage components. Thus, CSPs need to design a scalable, redundant, and efficient log collection platforms. In addition to high velocity, Cloud logs are generated by various platforms and applications that translate to a high degree of variety in collected logs [49]. An efficient Cloud monitoring platform should be able to support various types of Cloud logs.

Peculiarly, Cloud generated logs have fluctuating value. That is, during the normal operation of a Cloud platform (when components are running smoothly) logs do not have operational value except for the capacity planning. However, as soon as a technical issue arises, logs become extremely important, and their value increases.

In addition to a high volume of generated logs by the primary infrastructure components, all CSPs, such as Amazon, Google, IBM, and Microsoft rely on highly redundant hardware and software infrastructure. That is for each router, each firewall, and each server, there are many instances that are running or stand-by. Logs generated by redundant devices need to be filtered and only stored if they bring technical or business value. Hence, Cloud monitoring tools should have the ability to quickly filter the log and make a decision about the storage of data.

Besides redundant infrastructure, unique characteristics and features of Cloud computing environments, such as elasticity and auto-scaling, cause significant challenges for the Cloud monitoring tools. The number and nature of deployed resources in a pre-Cloud environment are mainly static; however, the elasticity of the Cloud results in a dynamic environment, in which, additional resources are dynamically added or removed. Hence, causing significant challenges for the monitoring tools that are designed for static environments [63]. As the Cloud offerings have to scale elastically [64], efforts have been made to build monitoring tools using multi-tier and peer-to-peer architecture, making the tools more resilient to elasticity compared to conventional monitoring systems [64].

Implementation of a complete monitoring system requires full access to the components that will be monitored [52]. Nguyen et al. [52] indicated that only Cloud providers have such level of access to Cloud resources. Therefore, most of the Cloud monitoring solutions are built by Cloud providers. In contrast, Cloud consumers require full details of monitoring data and need a way to verify the monitoring details that are provided by the Cloud provider. To address this issue, Nguyen et al. [52] combined role-based monitoring templates with agent-based monitoring and used an event processing engine to refine the collected data and to provide a trustworthy and holistic monitoring solution. Due to the challenges mentioned above, almost all Cloud monitoring solutions are implemented by the CSPs. Amazon CloudWatch [1], Azure CloudMonix [22], and IBM Cloud Event Management [6] are among such monitoring platforms. The list of issues discussed in this section is not exhaustive, for additional ones see [56].

## 2.2 Challenges of storing and analyzing Cloud-generated logs

The storage requirement for Cloud-generated logs (discussed above) calls for innovative expansion of Cloud storage features. Vernik et al. [61] suggest the use of federated storage to increase the capacity and performance of Cloud storage to accommodate demanding storage tasks (such as log storage). By nature, logs are redundant, and systems continuously generate and write logs even if the values for monitored metrics do not change. Hence, to preserve space required to store logs, Anwar et al. suggest to avoid storing repetitive values, leading to reduction of the size of stored data by up to 80% [30].

The high volume and velocity of generated monitoring data pose various challenges for monitoring systems. Not only the storage of such data is a challenge, but also the data processing portion is computationally expensive [49, 64] and sometimes infeasible [32, 50]. Hence, storage and processing the data generated by Cloud monitoring platforms is one of the key challenges of Cloud monitoring [64]. Using the same example provided in Section 1, consider a Cloud log storage platform that stores about 2.35 TB data per day. Simply querying the data (that is reading it and loading it to any application) by itself is a major challenge. Some researchers have proposed solutions [47, 62] based on the combination of Big Data storage and processing tools, such as HDFS [11] and Spark [5], with Cloud monitoring solutions. The data can then be passed to anomaly detecting techniques, such as [29, 32, 35, 41].

## 3 OBJECTIVES

As discussed above, Cloud computing issues, especially infrastructure issues, should be detected and fixed as fast as possible. However, this means that issues should be detected in near-real-time so that the IBM DevOps team can be informed about the issue and start their troubleshooting. As a manual observation of logs, at least at the scale of Cloud logs is impossible, our **primary objective** is to design, and implement a platform that can be used for collecting, storing, and analyzing Cloud-generated logs. The platform should have non-functional features, such as scalability, reliability, and redundancy (which we will elaborate on in Section 4.1).

Our **secondary objective** is to provide a solution that helps CSPs to reduce their operational cost by detecting the right problem at the right time. Obviously, the reliability and performance of a Cloud platform directly translate to customer satisfaction and perhaps returning customers. Such a system can be used to save time for IBM DevOps teams, minimize human error, and detect issues as they unfold. In other words, the automated log analysis results in fast and prompt detection of unusual behaviour and alerts teams to take care of issues promptly. As a representative use-case for this objective, we will analyze alerts emitted by components of an IBM Cloud service (which will remain anonymous to preserve confidentiality) and categorize these alerts into true and false ones, as will be discussed in Section 5.1.

We achieve these objectives by implementing a prototype of the monitoring system for IBM Cloud components using publicly available IBM Cloud services. We report our approach and findings below.

## 4 METHODOLOGY

In this section, we provide technical details related to the proposed solution. We indicate the modules that are used, their order, the data flow, and we conclude this section by reviewing the architecture of the proposed solution. The proposed solution has the capability of listening or retrieving data from various Cloud components and offers scalable storage and monitoring infrastructure for log analysis. By using the proposed solution, a massive amount of logs can be collected, transformed to the desired format, and stored in Cloud data storage platforms so they can be fed to analytical tool for further data analysis. Before we dive into details of our proposed solution, let us elaborate on the desired characteristics of a Cloud monitoring platform based on the content provided in Sections 1 and 2.

### 4.1 Desired characteristics

Here we list the desirable key characteristics of a Cloud-scale monitoring system.

*Scalability*: A platform that can monitor several metrics for each of the components in a Cloud platform distributed among several data centers needs to be scalable. The number of resources that needs to be monitored constantly increases and therefore, only a scalable monitoring platform can be used in Cloud environments.

*Elasticity*: A platform that should automatically scale up or down with the intensity of the incoming log records.

*Reliability and Stability*: A monitoring platform is mainly implemented to ensure that all other components of a Cloud platform is operating normally. As collection of data and monitoring of Cloud platform are continuous activities and require 24x7 operation, Cloud platforms should be reliable and resilient to component failure.

*Capacity*: The scale of generated logs requires a monitoring system with elastic capacity. That is, the size of the collected logs continues to grow and so does the required space to store them.

*Support of various log formats*: Cloud platforms consist of several different types of components. The logs generated by various components (e.g., servers, routers, and software platforms) are of different types and a Cloud monitoring platform should be able to collect and process various types of logs.

*Interconnection Feasibility*: As Cloud providers continue to add new services, it is very important that the existing monitoring platform can keep up with new demands.

### 4.2 Architecture

We are basing this solution on the 7-layered architecture for processing online and offline data described in details in [42, 43]. This architecture allows us to use micro-services and publish-subscribe software architecture and offers a good balance between scalability and maintainability due to high cohesion and low coupling of the solution. Furthermore, asynchronous communication between the layers makes the 7-layered architecture a building block for a general architecture for processing streaming data.

The detailed description of the 7-layered architecture is given in [42, 43]. Below we provide a summary of this architecture, depicted graphically in Figure 1. Microservices in odd layers communicate with each other via topics of a publish-subscribe infrastructure in even layers. A microservice in Layer 1 ingests data (e.g., log records and metrics' observations) from external sources and converts them to a unified format that is recognized by the subsequent layers of the architecture; it then publishes the converted data in a message to a topic in Layer 2. A microservice in Layer 3, subscribed to a topic in Layer 2, receives a converted message from the Layer 1 and decides what to do with it: either publish a message to topics of interest in Layer 4 or discard the message (if it is not of interest to any model). A microservice in Layer 5, subscribed to topics in Layer 4, aggregates received messages, enriches them with historical data (received from the persistent storage), and transforms the resulting data in the form required by the models in Layer 7. The resulting transformed and enriched data are passed, via topics in Layer 6, to models residing in microservices of Layer 7. The output of the models (e.g., a label deeming given observation anomalous or not) is passed to external services (e.g., a system, such as PagerDuty [23] or Slack [26], notifying Operations team about anomalies).

The microservices interact with the persistent storage for storing and accessing historical data and trained models. Moreover, it is often used for caching batches of data, as the size of a message passed via publish-subscribe software is limited [43]. For example, at the time of writing, Amazon Kinesis and IBM Event Stream maximum message size is $\approx$ 1 MB [16, 21]. In practice, batched input data, historical datasets prepared in Layer 5 for retraining models, and the trained models (created and reused in Layer 7) often exceed this limit. One can split the large data into chunks and pass these chunks in separate messages via publish-subscribe. However, given that messages are passed asynchronously, aggregating them on the receiving end becomes problematic. Instead, 7-layered architecture recommends to persist the data (e.g., to object store or database) and pass a message with a pointer to the stored data via publish-subscribe [43].

### 4.3 Inputs: pulling and pushing

Many Cloud components (such as servers, routers, and software platforms) produce the logs which are then collected and stored in various form of databases [54]. Both relational and non-relational databases (such as InfluxDB [18], ApacheHbase [2], and Firebase [10]) are used to store, either temporarily or permanently, collected logs of Cloud platforms [65]. Irrespective of the type of databases that stores the Cloud logs, our platform needs to have the capability of pulling (retrieving data) from such data sources. The *pulling* happens either by using the database-provided APIs that expose an interface to retrieve data or through database-provided software development kits and libraries.

Similarly, many Cloud monitoring tools can be integrated with other 3rd-party platforms by calling an API that is exposed on the 3rd-party side. This *pushing* mechanism is often predefined and requires little to no changes in the current structure of the monitoring platform. Hence, it is one of the most common practices of system integration.

As it was reviewed above, a scalable and useful Cloud monitoring tool need to have both 'pushing' and 'pulling' capabilities so that other components, namely the components that either generate logs or hold logs, can easily interact with the proposed tool. Figure 2 depicts the interconnection between our proposed solution and Cloud-generated logs.

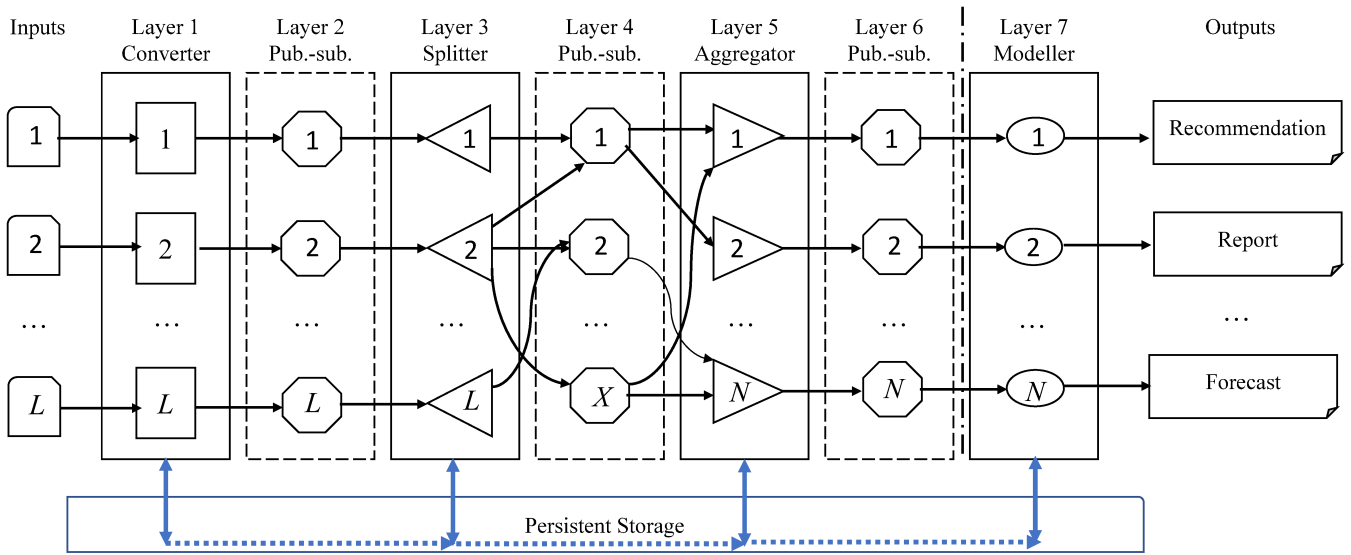

**Figure 1: A diagram of the 7-layered architecture with $L$ input sources, $N$ models, and $X$ topics of interest. Dashed lines represent publish-subscribe layers. Vertical dash-dotted line separates data preparation layers from the analytics layer. Arrows denote flow of data. Blue arrows between Layers 1, 3, 5, 7 and persistent storage reflect potential communication between micro-services (in a given layer) and persistent storage. Adapted from [43].**

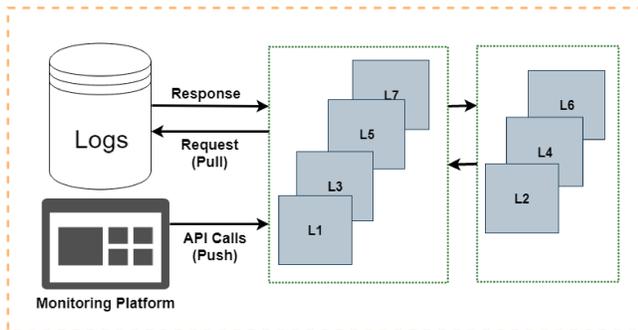

**Figure 2: Interconnections between the proposed solution and its inputs.**

Layer 1 of the 7-layered architecture is exposed to the outside world for data collection. A RESTful API [38] is designed and implemented, and logs can be pushed to this API.

## 4.4 Implementation

As discussed in Section 3, in our prototype of the monitoring system, we leveraged public IBM Cloud services to implement the 7-layered architecture. The relations between the services are represented graphically in Figure 3, details are given below.

### 4.4.1 Publish-subscribe.
We chose public multi-tenant offering of IBM Cloud Event Streams [9] (based on the Apache Kafka [3] software which implements publish-subscribe pattern). The offering if fully managed, highly available, and is charged by the number of topics and the amount of data passing through them.

### 4.4.2 Microservices.
We could have implemented and manage microservices using managed container offering (namely, IBM Cloud Kubernetes Service [14]). However, we decided to leverage fully managed function-as-a-service (FaaS) offering of IBM Cloud Functions [7] (which is implemented based on Apache OpenWhisk [4]). Given that the 7-layered architecture is aimed at the Data Science pipeline, which is typically stateless, this was a natural decision. Solutions on FaaS scale up and down trivially, as the FaaS platform 'fires up' individual instances of a Function to process incoming messages: the more messages are in the pipeline, the more messages will be running concurrently. It was also attractive economically: FaaS are billed based on the amount of time a Function is running (proportional to the number of CPUs and memory requested to each Function); thus, we do not have to pay for idle resources.

We implement the Functions (from here on we will use the term 'Function' and 'microservice in the 7-layered architecture' interchangeably) in Java or Python. Given that the microservices are communicating with each other asynchronously via topics in the IBM Cloud Event Streams, the usage of multiple languages was not an issue.

IBM Cloud Functions have multiple types of triggers to start a Function. Our prototype receives incoming data via RESTful API [12]. IBM Cloud has readily available service to deploy the API and authorize access to it [13]. Triggers are set up to call a corresponding Function in Layer 1 for every POST request coming through the API.

IBM Cloud Functions has a specific type of triggers for the Event Streams. Such trigger listens to an Even Streams' topics and starts an instance of a Function when a message (or a group of messages) are published to this topic. The trigger then passes these messages

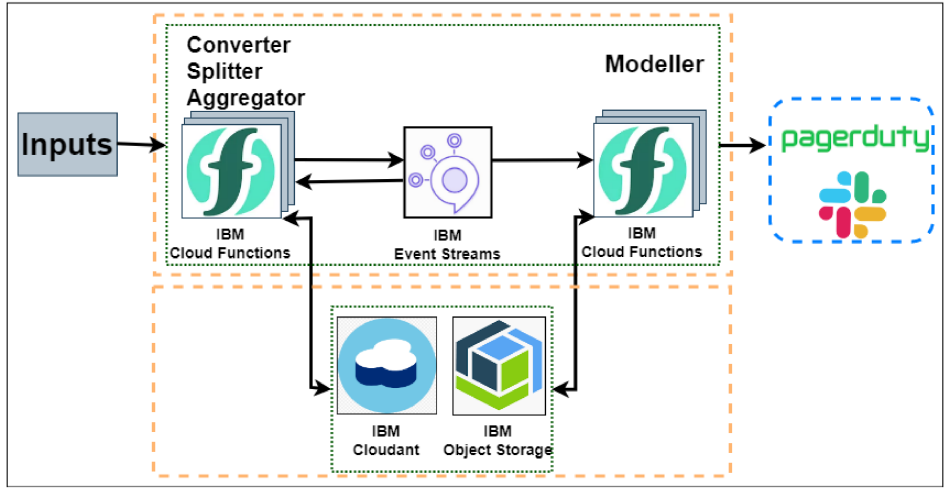

Figure 3: Architecture of the proposed solution.

to the started Function. This is how we activate the Functions in Layers 3, 5, and 7.

Some of the models are retrained periodically rather than online; for those models, we set up timer-based triggers that retrain the models (by firing specialized Functions in Layer 5 periodically).

*4.4.3 Persistent storage.* For persistent storage, we use two services. The first one is fully-managed JSON document store IBM Cloudant [15], which elastically scales throughput as well as storage. However, a single document, stored in Cloudant, cannot be larger than 1 MB [8], which implies that we cannot store large transformed data frames (produced in Layer 5) and trained models (used in Layer 7). For these items, we leverage IBM Cloud Object Storage [15], another fully-managed and scalable service.

## 5 RESULTS AND DISCUSSION

### 5.1 Practical test

To assess the feasibility of the prototype, we put it to a practical test. A DevOps team of the IBM Cloud services implemented a monitoring solution for approximately 2,000 deployed instances of this service. The solution checks values of a set of metrics and emits an alert if a metric violates a predefined threshold. The alerts are then passed to PagerDuty for DevOps to monitor.

Predefined thresholds may lead to the generation of a large number of false alarms, which leads to a waste of DevOps resources [56]. To address this, we built a machine learning model (based on the Random Forest algorithm [33, 60]) that does further analysis of these alerts and classifies them into false and true alerts to reduce the load on the DevOps team. We then used our solution to deploy the model in production as follows.

A PagerDuty web hook [25] is created to POST to our RESTfull API every change made to the PagerDuty incident tickets of the DevOps team. In a Function of Layer 1, we coded up the converter of PagerDuty JSON records into our internal format. The Function passes the records to Layer 3 via topic in Layer 2.

Layer 3 Function saves all messages to Cloudant (as they are used to calibrate the model). The Function then submits to a topic in Layer 4 only the messages that relate to newly created accidents (which we would like to classify with our model as true or false alerts). A Function in Layer 5 reads these messages and enriches the data with historical information (extracted from the Cloudant database). It then passes the message (via topic in Layer 6) for prediction to a Function in Layer 7. The Function in Layer 7 loads pre-trained model from the Object Store, performs classification of a new incident and feeds the results of classification back to the DevOps team by posting a message to a Slack channel monitored by the team.

We recalibrate the model periodically. The recalibration is implemented in two Functions: a Function in Layer 5 prepares a historical dataset for training; the Function then saves the historical dataset to Object Store and passes a pointer to the dataset via Layer 6 to a Function in Layer 7 (which, in turn, retrains the model). The newly trained model is then saved to Object Store (overwriting the previous version). The model retrains on approximately 2500 features (extracted from the historical data). The retraining process takes approximately 5 minutes and is done in parallel with running the predictions.

The predictive performance of the model is adequate[3], as shown by the receiver operating characteristic (ROC) curve [37] in Figure 4. The DevOps team can then prioritize the alerts based on the probability that a particular alert is a true alert, as reported by our model.

The prototype of our solution was running for a month, analyzing alerts from the 2,000 deployments discussed above. We computed the duration of the time from a message arriving into our system to the message leaving the system (depending on the type of content in the message, we either produce a report in Layer 7 or save it to persistent storage in Layer 3). The statistics of the timing data are shown in Table 1 and Figure 5.

---
[3] We do not imply that this model yields the best results for the task at hand. Instead, it serves as an illustration of a typical use-case.

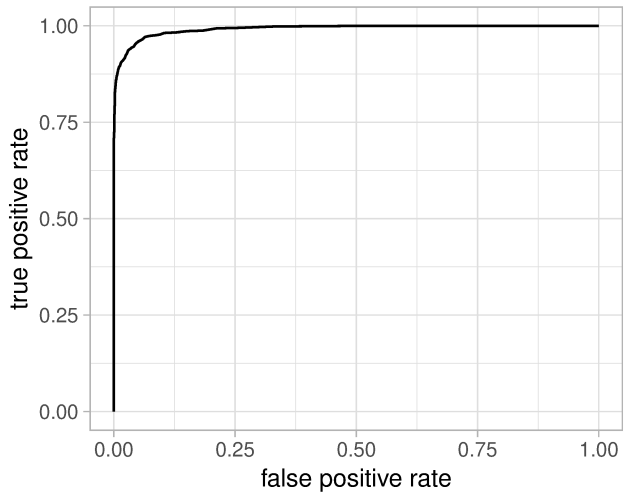

Figure 4: A receiver operating characteristic curve of the prediction model from one of the retrainings.

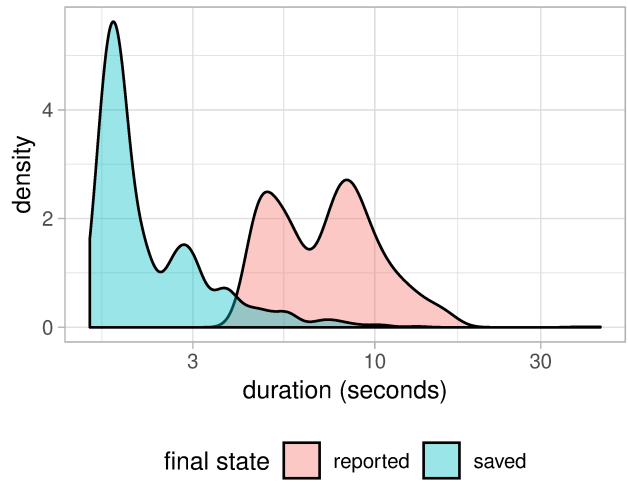

Figure 5: Timing of the prototype of the monitoring system. The figure shows distributions of the duration of time from a message entering the prototype to the message leaving the prototype. If a message left the system in Layer 3 (by been saved to persistent storage), it is deemed 'saved'. If a message left the system in Layer 7 (via a produced report), it is deemed 'reported'.

Table 1: Summary statistics of the timing data.

| Type | Min. | 1st Qu. | Median | Mean | 3rd Qu. | Max. |
| --- | --- | --- | --- | --- | --- | --- |
| Reported | 4.1 | 5.3 | 7.5 | 7.7 | 9.2 | 17.0 |
| Saved | 1.5 | 1.7 | 1.9 | 2.5 | 2.7 | 44.5 |

As we can see, on average it takes 8 seconds for the whole lifecycle (from receiving the message to emitting the report); at worst, it takes 17 seconds. Notice that the right tail of the distribution is heavy: in fact, 75% of the reports (3rd Quartile) complete in 9 seconds or less.

As for historical data (saved in Layer 3), it takes on average 3 seconds (and, at worst, 45 seconds) to reach the persistent storage. The maximum value of 45 seconds is an outlier: 99.87% of messages were saved in 15 seconds or less.

## 5.2 Does our implementation have desired characteristics?

We discussed the desired characteristics of a Cloud-scale monitoring system in Section 4.1. Here we provide details as to whether our proposed solution matches these characteristics.

*Scalability*: We have observed that our implementation is able to handle from 0 to 120 alerts per minute[4]. The components of our prototype seamlessly scale up and down with the load, without affecting the processing time of the messages.

The 7-layered architecture spreads atomic tasks between the Functions. For each new message that is submitted via API to Layer 1, IBM Cloud Functions and their triggers, instantiate new and isolated instances of Functions. Each Function independently processes a log record and sends the results to the next layer by using reliable message delivery of IBM Event Streams. At the IBM Event Streams side, the concept of the client-specific topic is implemented. That is, each generator of logs can have a dedicated topic so that listeners and publishers can interact with it by sending or receiving messages. Topics are highly scalable as one may configure multiple topic clusters. Evidently, the platform is scalable for message processing (using the Functions), for delivering the messages between the functions, and for persisting the messages and associated metadata (using Cloudant and Object Store).

*Elasticity*: The behaviour described in the previous two paragraphs contributes to elasticity as well: the Functions are fired upon demand, and we pay only for the CPU and memory resources consumed during their execution. Multi-tenant publish-subscribe and persistent storage services grow and shrink elastically with the workload, and we are charged for the services based on their intensity.

*Reliability and Stability*: The proposed solution is highly reliable and stable as the core functionalities are divided among various layers, and therefore, the operation of the proposed solution is decentralized. Moreover, for each of the IBM Event Streams, as well as the IBM Cloud Functions, enterprise-grade categories exist that come with even higher availability and redundancy settings.

*Capacity*: IBM Cloudant standard edition comes with 20 GB of storage, but additional storage can be added on a pay-as-you-grow basis [24]. If required, space savings can be achieved by implementing advanced log storage techniques, such as deduplication [30] and aggregation of log records.

*Support for various log formats*: As long as a log record is parsable (in a loose sense of the term) and the record's format is known, one can implement a converter in Level 1 to standardize the format. Thus, this characteristic is achievable.

---

[4]Typically, intense arrival of alerts indicates an issue with supporting infrastructure rather than the actual instances of the service been monitored.

*Interconnection feasibility*: given that a new service has a pull or push capability to deliver log records, it can be easily integrated with the proposed solution.

In summary, the proposed solution relies on various layers, and every single layer can be equipped by exception handling logic to ensure a robust and stable Cloud-generate log storage system. The use of publish-subscribe as an internal messaging bus ensures the capacity and scalability that is needed for transactions among layers. Moreover, the use of serverless architecture significantly increases technical and financial feasibility as Functions are only called when there are data to process.

### 5.3 Lessons learnt

*5.3.1 Processing time.* As we have seen from timing analysis in Section 5.1, in rare cases, the prototype takes longer than usual to process a message, although the processing, even in the worst-case scenario, did not exceed 1 minute. This can be explained by the fact that the prototype is deployed in the multi-tenant environment and has to 'compete' for resources with workloads of other tenants. This timing was acceptable for our use-case. If one requires lower processing time, a standalone deployment of IBM services used to implement the prototype (which are still hosted and maintained by IBM staff) may help to achieve this goal. This may be more expensive for a smaller deployment and more economical for a larger one.

*5.3.2 Ease of testing.* We unit test the code of each Function in isolation. However, when it comes to integration and system testing, it is desirable to do it right on the IBM Cloud. IBM Cloud Functions service has the option to create Functions in different namespaces, making it easy to create separate development, testing, staging, and production environments.

But what data should we feed to the Functions of Layer 1? In the case of the development and testing environments, one can leverage predefined synthetic inputs. However, for staging environments (and even for higher-order testing) it is more prudent to use live data. It is not economically feasible to duplicate every API request coming into the production environment to the test and staging environments; instead, a fraction of the production requests should be cloned to these environments.

This can be done by putting an extra layer in front of the Layer 1 (let us call it Layer 0). The Functions in Layer 0 are sending a subset of production data to corresponding Functions in Layer 1 of the staging and testing environments (e.g., every 100th message goes to the staging environment), while every single message is sent to the Functions in Layer 1 of the production environment. The sending from Layer 0 to Layer 1 can be done either directly (if we are confident in the quality of our code) or via publish-subscribe topics (which is a safer choice).

*5.3.3 Resiliency to defects.* IBM Cloud Event Streams (and Apache Kafka under the hood) has a sophisticated mechanism for committing an offset in a stream of data. When a batch of messages is published to a topic, one can rollback the whole batch if one or more messages in the batch failed to be published. IBM Cloud Functions can access this functionality: when a Function posts the message to Event Streams topic, it does so by calling Kafka Producer API [20].

Similarly, when a consumer (i.e., the Function) reads a message from a topic to which it is subscribed to, in the case of failure of processing the message, it may report it back to the Event Streams engine, and the message will be retained for processing by a different consumer.

This feature is handy if the consumer's code has a defect which prevents it from processing the message. A developer can look at the failure, fix the error, and then re-process the message without risk of losing the message's payload.

However, to leverage this functionality, the consumer's code has to be able to access the Kafka Consumer API [19]. By design, the Functions are triggered by a specialized Event Streams trigger that fires a Function every time a message comes into the topic. This trigger operates in a 'fire and forget' mode: it takes the message, passes it to a Function, and tells the Event Stream engine that the message was 'processed successfully', independent of the outcome of the Function's execution. Once the message is marked as 'processed successfully', the Event Streams engine will delete the message (even if in reality the Function has failed). There are three potential ways to rectify this issue.

(1) Write bug-free code, which is desirable but not realistic as to err is human.
(2) Move the code from IBM Cloud Event Streams service to IBM Cloud Kubernetes Service. The Kubernetes microservices can call Kafka Consumer API directly. From the coding perspective, such migration from the Functions to Kubernetes would be relatively inexpensive, as we will simply need to add proper Kafka Consumer API to the code of every Function. However, the runtime costs may increase, as the microservices will have to listen to the Event Streams topics constantly.
(3) Enhance functionality of the Event Streams trigger by adding an ability to listen to the outcome of the Function, so that a message will not be marked as 'processed successfully' in the case of the Function's failure. We hope that developers of Event Streams triggers will add this functionality in the future.

*5.3.4 Usability issues.* A minor usability issue relates to the set up of a periodic trigger (used to start recalibration of a model): once the trigger is created, its schedule cannot be altered. When we needed to modify the schedule, we had to delete an existing trigger and create a new one. This is a minor inconvenience and, hopefully, will be rectified in the future by the developers of the triggers.

## 6 CONCLUSION

In this paper, we report our experience of creating a prototype of the scalable and resilient platform (based on the 7-layered architecture), for monitoring and analysis of logs emitted by components of IBM Cloud services. To implement the prototype, we leveraged public IBM Cloud services, hence the dogfooding.

The prototype was tested on the production data. It showed good scalability (been able to process up to 120 requests per minute) and responsiveness (reporting the results of the analysis in ≤ 17 seconds).

In the future, we are planning to widen the usage of our platform, by monitoring additional IBM Cloud services, and to integrate into the platform our immutable blockchain-assisted log storage.

## ACKNOWLEDGMENTS

The authors would like to thank IBM Centre for Advanced Studies for their generous support.

This research is funded in part by IBM CAS Project No. 1046 and NSERC Discovery Grant No. RGPIN-2015-06075.